\documentclass{article}
\usepackage{PRIMEarxiv}
\usepackage{amsmath}
\usepackage[utf8]{inputenc} 
\usepackage[T1]{fontenc}    
\usepackage{hyperref}       
\usepackage{url}            
\usepackage{booktabs}       
\usepackage{amsfonts}       
\usepackage{nicefrac}       
\usepackage{microtype}      
\usepackage{lipsum}
\usepackage{fancyhdr}       
\usepackage{graphicx}       
\usepackage{xcolor}
\graphicspath{{media/}}     
\usepackage{multicol}
\usepackage{longtable}
\usepackage{chemformula}
\title{The diaspora model for human migration}

\author{
  Rafael Prieto-Curiel\\
  Complexity Science Hub \\ Josefstaedter Strasse 39 \\ 1080 Vienna, Austria \\
  \texttt{prieto-curiel@csh.ac.at} \\
   \And
  Ola Ali \\
  Complexity Science Hub \\
  \texttt{ola@csh.ac.at} \\
  \And
  Elma Dervic \\
  Complexity Science Hub\\
  \texttt{dervic@csh.ac.at} \\
  \And 
  Fariba Karimi \\
Vienna University of Technology (TU Wien)\\
Graz University of Technology (TU Graz)\\
Complexity Science Hub \\
  \texttt{karimi@csh.ac.at} \\
  \And 
  Elisa Omodei \\
  Central European University \\ Vienna, Austria\\
  \texttt{omodeie@ceu.edu }\\
\And 
Rainer Stütz \\
Complexity Science Hub \\
\texttt{stuetz@csh.ac.at}\\
\And 
Georg Heiler \\
Complexity Science Hub \\
\texttt{heiler@csh.ac.at}\\
\And 
Yurij Holovatch \\
Institute for Condensed Matter Physics \\ National Academy of Sciences of Ukraine, Lviv, Ukraine. 
\\ L4 Collaboration (Leipzig-Lorraine-Lviv-Coventry)\\ 
Centre for Fluid and Complex Systems, Coventry University \\
Complexity Science Hub \\
\texttt{hol@icmp.lviv.ua}
}              

\begin{document}
\maketitle

\begin{abstract}

Migration's impact spans various social dimensions, including demography, sustainability, politics, economy and gender disparities. Yet, the decision-making process behind migrants choosing their destination remains elusive. Existing models primarily rely on population size and travel distance to explain flow fluctuations, overlooking significant population heterogeneities. Paradoxically, migrants often travel long distances and to smaller destinations if their diaspora is present in those locations. To address this gap, we propose the diaspora model of migration, incorporating intensity (the number of people moving to a country) and assortativity (the destination within the country). Our model considers only the existing diaspora sizes in the destination country, influencing the probability of migrants selecting a specific residence. Despite its simplicity, our model accurately reproduces the observed stable flow and distribution of migration in Austria (postal code level) and US metropolitan areas, yielding precise estimates of migrant inflow at various geographic scales. Given the increase in international migrations due to recent natural and societal crises, this study enlightens our understanding of migration flow heterogeneities, helping design more inclusive, integrated cities.

\end{abstract}

\section{Introduction}

{
Births, deaths and migration are the most relevant demographic components of population change, but migration is the most difficult to quantify, model and forecast \cite{UNReportMigration}. Mass movements of people change the spatial distribution of the population and explain why some places grow faster than others \cite{bettencourt2020demography, verbavatz2020growth}. The size of cities and hierarchy are heavily affected by migration \cite{geyer1996expanding, ABMCitySize, bell2015internal, PopGrowth, WorldMigrationReport2015}. Today, international migrants would form the fourth largest country in the world, and approximately 1\% of the World's GDP is sent as international remittances \cite{UNDesa, WorldBankRemitsKnomad}. Migration is a selective process that tends to attract young and highly skilled people into large cities, increasing the burden of human capital flight but easing economic disparities across borders \cite{koser2009migration, keuschnigg2019urban, de2017selection}. It is a core strategy for coping with unemployment, violence, or disasters \cite{BookClimateChangeMigration, DisasterMigrationBangladesh, ConflictMigrationSaharanAfrica, AgeOfMigration, kaczan2020impact}. Migration eases the pressure of an ageing population and alters the gender imbalance \cite{WorldBankMigration}. Accurately predicting the number of individuals who will relocate and their precise destinations holds significant implications for resource allocation, planning, and hosting strategies. Yet, most migration models have many critical weaknesses. Migration models often focus on where they will go but ignore a more critical question: how many will move. Also, they try to capture the attractiveness of a destination based on population size, so they tend to fail drastically at small spatial units, such as neighbourhoods. Finally, they distinguish migrants only by location on a map, so they fail when comparing many origins. Here, we construct a migration model with the aim of forecasting that distinguishes between the nationality of migrants, and that works accurately at the neighbourhood level.
}

{
Critical repeated patterns emerge when studying thousands of migrations. For example, most migration is across short distances, between nearby areas \cite{MigrationLaws1885, DistanceOnMigration, TheoriesOfMigration, barbosa2018human}. Also, large destinations attract disproportionally more migrants than smaller ones \cite{philbrik1973short, ScalingMigrationRPC}. The gravity model for human migration has been extensively used to capture both the impact of size and of distance simultaneously \cite{Gravity, GravityModel, anderson2003gravity}. However, the gravity model has many limitations, including parameter dependence, issues with estimation, and bias, among others \cite{GravityPanelData, ZIPMigration, beyer2022gravity}. Extensions of the gravity model have been constructed by looking at job opportunities, the distribution of points of interest, the road network, and other geographical attributes \cite{BarabasiMigration, simini2021deep, kluge2021evaluation, liu2020universal, alis2021generalized, masucci2013gravity, hong2019gravity}. The main drawback of these models is that they ignore the vast heterogeneity of migrants and distinguish individuals only by their location. As long as two individuals are at a similar distance to some destination, the gravity model (and its extensions) will give identical migration estimates. Thus, for these types of models, what makes a person different from others is only their location. Yet, we observe that migration is a much more complex process that depends on many more aspects. The reasons why a person leaves their home in Nicaragua are not similar to a person leaving their home in Italy or Senegal, and the same applies to destinations. Silicon Valley attracts a different kind of people than Las Vegas or Hollywood. Migration's push and pull factors do not apply equally to all people \cite{lee1966theory, PushPullTobler}. At a microscopic level, a similar process observed with residential selection profoundly impacts society, where mild preferences for living with similar neighbours may cause severe segregation and create politically homogeneous neighbourhoods \cite{SchellingSegregation, clark2008understanding, motyl2014ideological, bishop2009big}. Likewise, if migrants prefer to move to some community, they will naturally form a significant diaspora in the new destination. However, models do not capture the spatial sorting of individuals.
}

{
Various reasons explain why people with different backgrounds are attracted to distinct places. For example, an expensive neighbourhood receives wealthier people, students move to university towns, startups attract engineers, or cities such as Los Angeles attract artists. While numerous reasons could explain why an individual migrates to a new destination, we observe surprising regularities and predictability in migration flows. One explanation is related to the flow of information. Individuals are more attracted to places where they have more information, mainly acquired from their social networks. In this case, early migrants reduce uncertainty and provide adequate information for late arrivals, creating a self-reinforcing mechanism \cite{massey1999dynamics, DistanceOnMigration, SimonE7914, zavodny1997welfare, nedelkoska2021role}. Most people move to places where they have pre-existing ties \cite{RootExodus}. This process relates to one of the most fundamental forces of our social life, namely homophily, the tendency to interact with similar others \cite{mcpherson2001birds}. Group identity based on race and ethnicity constructs leads to homophily and affinity between people \cite{tajfel2004social}. Therefore, homophily can directly or indirectly affect people's decision to migrate to a specific neighbourhood. A prime example of such affinity is migration and co-location due to strong same-race and same-ethnic dating or marriage preferences \cite{hitsch2010makes, jackson2019human}.  
}

{
This study examines the influence of homophilic preferences on international migration. We use the term \emph{diaspora} to refer to a group of people from one nationality living elsewhere. Consequently, we distinguish individuals not solely by their geographic location but, more significantly, by their country of origin. Thus, diaspora encompasses similarities in race, ethnicity, and more. We will see that the diaspora is a much more accurate explanation for modelling and predicting future migration beyond factors like distance or a city's population. Many social processes tend to be highly homophilic, and migration is no exception. 
}

{
Here, we construct a novel migration model based on the pull impact of the diaspora. Instead of looking at population size, travel distance, or points of interest, our model uses only the diaspora size. We analyse two migration scenarios. First, we look at population registers in Austria and explore arrivals to the country from other parts of the world (SM A). Second, we use the international arrivals to metropolitan areas in the US to show the pull mechanism of diasporas (SM D). We show that migration is a highly homophilic process. Opposing the principles of the gravity model, migrants travel long distances and go to small cities if there is a sizeable diaspora in the destination. We estimate that diasporas have a pulling impact, where 10,000 individuals will attract roughly 1204 new arrivals yearly in the case of Austria. We show that diaspora size can accurately explain migration even at the neighbourhood level. The diaspora model is more precise than the gravity model, holding for both international arrivals at the postal code level in Austria and the metropolitan area level in the USA. 
}

\section{Results}

\subsection*{Diaspora migration model}

{
Although there are many reasons why a migrant from some country chooses to move to some destination, similar reasons applied in the past to previous migrants from that country. Instead of observing and modelling the reasons, the principle we apply here is to look at how many people were already attracted to some destination and use it as a proxy to forecast future migrants (Figure \ref{ModelDescription}). For a given destination, we model migration with a hierarchical model, where the first component captures how many people will arrive, and the second is deciding where they will move to \cite{kery2010hierarchical}. The first component is the \emph{intensity} of the migration flow \cite{abel2014quantifying, bernard2017comparing, bell2015internal}. The intensity of arrivals from country $i$ is modelled with a homogeneous rate $\lambda_i$. Thus, the hosting country expects $\lambda_i$ migrants from $i$ daily. The second component is the \emph{assortativity}. Once a person decides to move, they choose $j$ destination among $\nu$ options in the hosting country (for example, metropolitan areas or neighbourhoods) with a Multinomial distribution. The probability that a person arriving from $i$ goes to $j$ is $\pi_{ij}$, with $\sum_j \pi_{ij}= 1$. Combining the two components ---intensity and assortativity--- the arrivals to destination $j$ have a homogeneous rate $\lambda_i \pi_{ij}$. In essence, we model first how many people will move to a country and then how they choose a specific destination (SM B). We show that arrivals to a country can effectively be modelled by a constant rate, which depends on the country of origin (see the Methods). 

\begin{figure}[ht!] \centering
\includegraphics[width=0.6\textwidth]{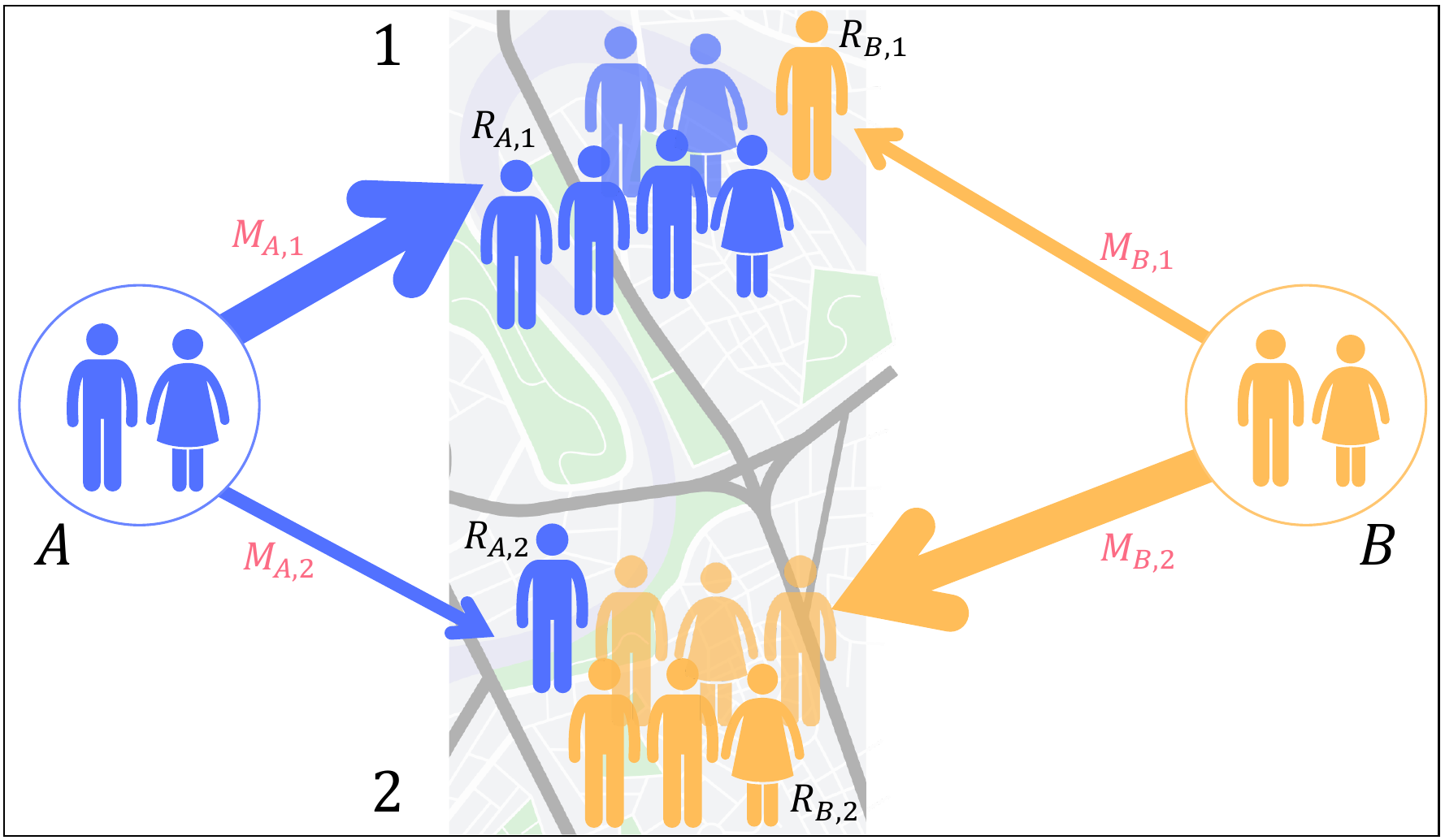}
\caption{\textbf{Illustration of the diaspora migration model.} We divide migration into two separate components: intensity (related to the arrival of migrants) and assortativity (related to where migrants decide to go). The diaspora model uses the size of the pre-existing population (depicted as the people on the map with different colours to represent people with different backgrounds) to estimate a steady inflow of migrants (represented by the arrow thickness) and their distribution across two regions (marked as 1 and 2 in the map).}
\label{ModelDescription}
\end{figure}
}

{
We capture the degree to which migration is a homophilic process by looking at the size of the diaspora (Figure \ref{ModelDescription}). We assume that $\lambda_i$ can be expressed as $\rho R_i$, where $R_i$ is the total diaspora from country $i$, so the arrival rate to the country depends only on the size of the diaspora and a \emph{pull rate} $\rho$ that applies equally to all countries. Then, we assume that the assortativity can be expressed as $\pi_{ij} \propto R_{ij}$, where $R_{ij}$ is the diaspora from country $i$ in destination $j$ (so $R_i = \sum_j R_{ij}$). Combining both assumptions, arrivals from country $i$ to location $j$ have a rate $\rho R_{ij}$. Consequently, the expected number of migrants from $i$ to destination $j$ for $t$ days, is 
\begin{equation} \label{DiasporaModel}
M_{ij}(t) = \rho R_{ij} t,
\end{equation}
reflecting that arrivals depend on the size of the diaspora (details in the Methods). Ignoring other demographic processes (births and deaths), the diaspora model results in the conservation of assortativity, meaning that the way migrants have been distributed in the past among neighbourhoods will be the observed pattern in the future. Thus, within the time window considered, assortativity is stable (SM A). 
}

\subsection*{Austrian migration dynamics}

{
Migration data is often scarce and requires long periods of observation to distinguish between migration and other types of mobility \cite{alexander2020combining, PrietoColombiaMobility}. Here, we use individualised data, which captures the primary residence of all foreign-born individuals in the country. Population registers capture all address changes and have become the primary source of migration data \cite{bell2015internal, poulain2013central, falkingham2016residential}. Population registers corresponding to all arrivals to Austria before a fixed date (December 2022) and 200 days later (labelled as ``arrivals'') are used to test the model. The data contains information regarding 1.46 million foreign-born individuals living in Austria, and it is used to determine the size of the diaspora of all countries at the postal code level, $R_{ij}$ (SM A). 
}

{
Arrivals to Austria are used to quantify the intensity and assortativity of the migration flow. In total, 111,244 individuals arrived in the country during the 200 observed days. The daily pull rate is around $\rho_{Aus} = 3.29\times10^{-4} $ per person (see the Methods). Hence, we expect one arrival for every $1/\rho_{Aus} \approx 3,031$ people, and the same applies to any diaspora and any destination considered (Figure \ref{Flows}). The diaspora model estimates the number of arrivals from any country to any destination at granular levels (data for each destination is shown by a disc in Figure \ref{Flows} B, C). It gives consistent estimates for geographic units (neighbourhoods) that can be aggregated to more extensive areas (such as cities or provinces). 

\begin{figure}[ht] \centering
\begin{center}
\includegraphics[width=0.7\textwidth]{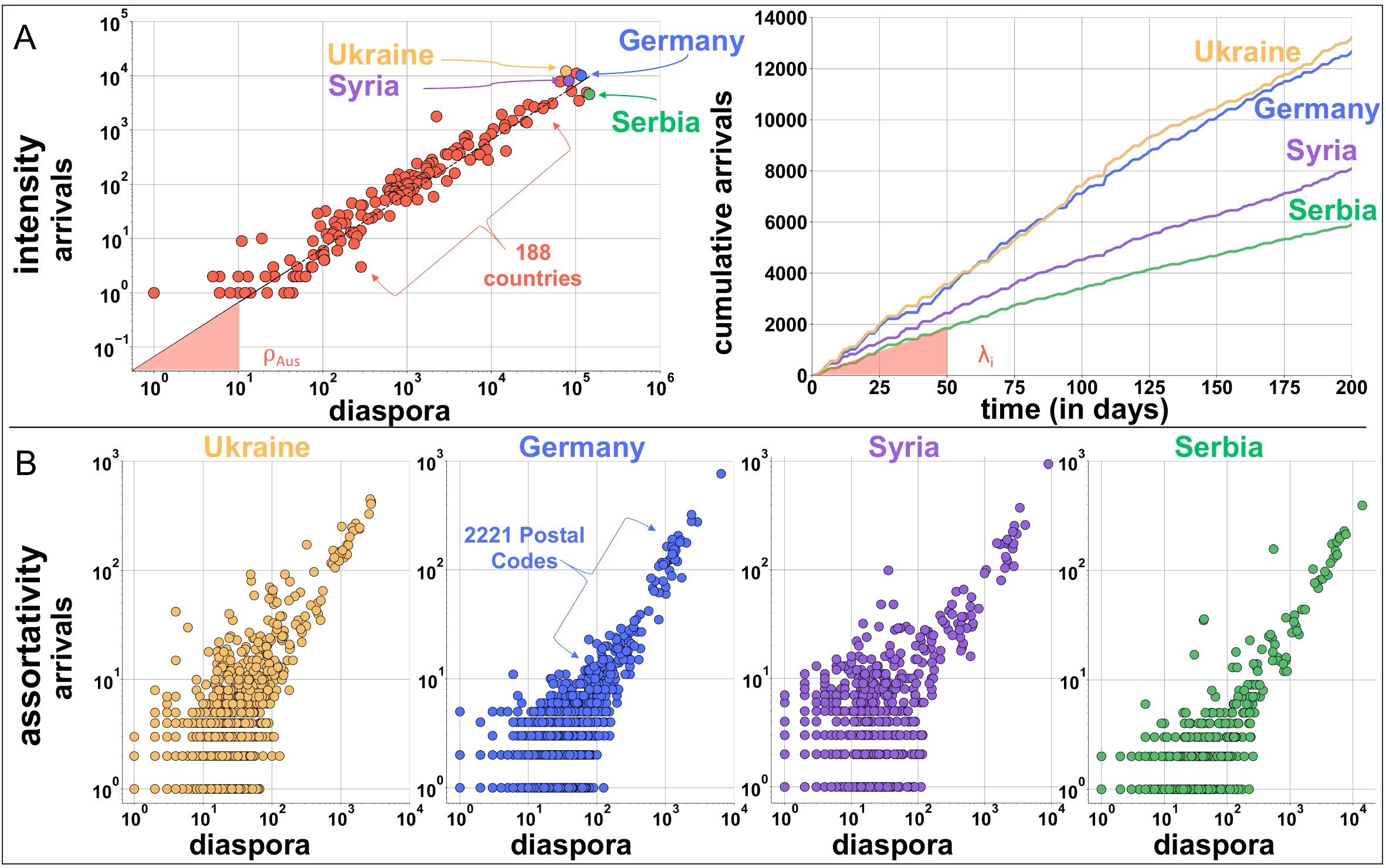}
\caption{\textbf{Observed intensity and assortativity of migrants.} A - (Left) the intensity of arrivals concerning the pre-existing diaspora size. Each point represents a country of origin, the diaspora size is the number of existing migrants from that country of origin in Austria before December 2022, and the arrivals is the number of new migrants observed after 200 days. The black dotted line is the daily pull rate of $\rho_{Aus}$. We highlight arrivals from Ukraine, Germany, Syria and Serbia because they belong to countries of origin with the biggest diaspora size in Austria and have different economic and political backgrounds and migration histories (SM A). (Right) observed cumulative arrival of the top diasporas within our observation period, with different arrival rates $\lambda_i$. B - Assortativity of migrants from the top diasporas. Each point represents a postal code in Austria (other diasporas in SM A).}
\label{Flows}
\end{center}
\end{figure}
}

{
To assess the predictive power of the diaspora model, we compare it with the gravity model for the top nationalities arriving in Austria (Figure \ref{ModelResults}). The mean square error of the gravity model is 2.85 times bigger than the diaspora model. The gravity model is particularly weak in differentiating nationalities but also for small geographical areas and does not offer a clear metric for predicting arrivals (SM C).   

\begin{figure}[ht] \centering
\begin{center}
\includegraphics[width=0.7\textwidth]{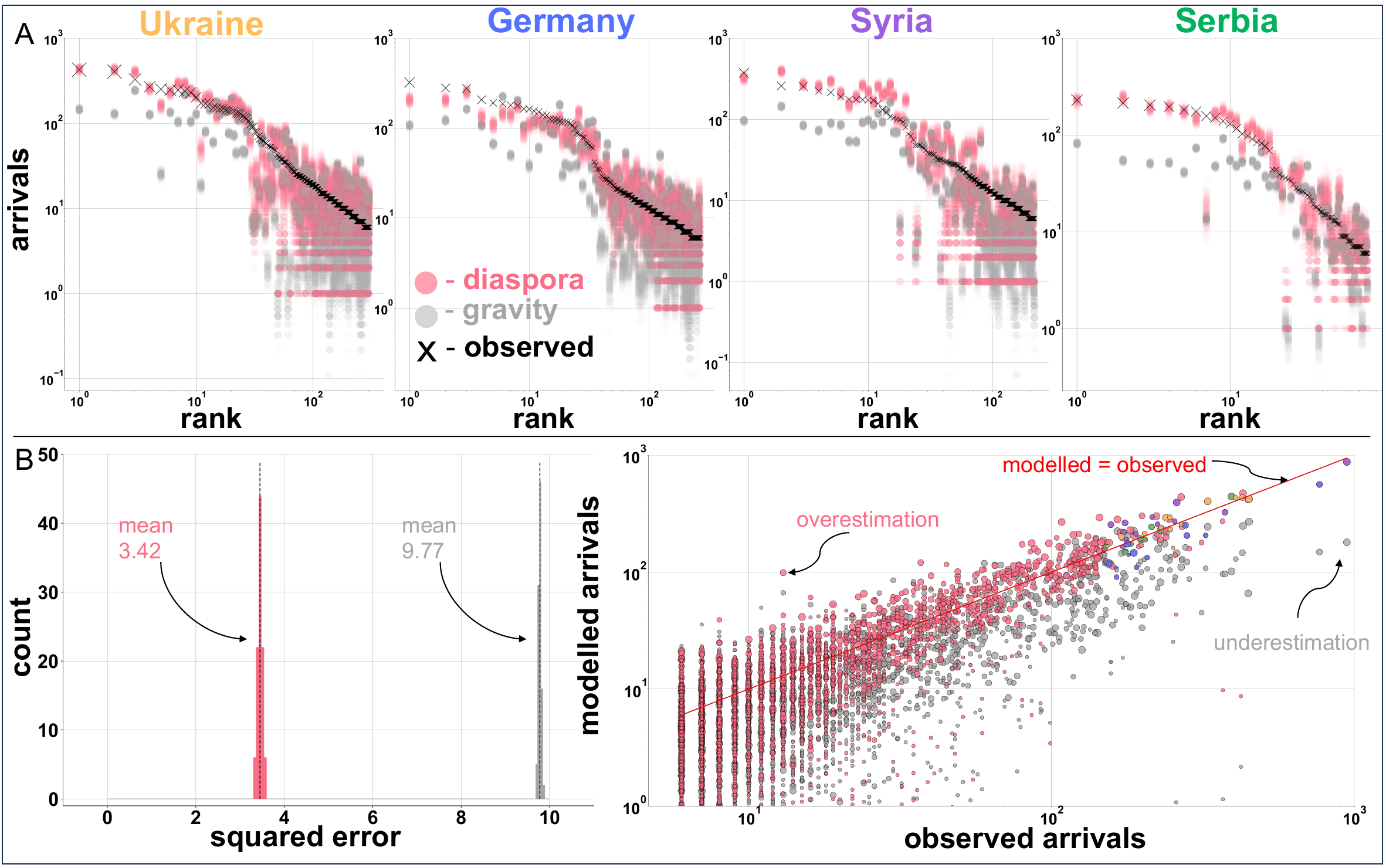}
\caption{\textbf{Model results and error estimation.} A - Results of the diaspora model for migration (pink) compared to the gravity model (grey) and the empirical observations (black crosses) over 100 runs. The horizontal axis is the observed number of migrations in each postal code, while the vertical axis is the number of observed migrants. We only show postal codes with arrivals above five during the observation period. B - The mean square error of our diaspora model (pink) vs. gravity (grey) over 100 runs. C - Modelled vs. observed arrivals to Austria. The disc size is proportional to the pre-existing population at the postal code } 
\label{ModelResults}
\end{center}
\end{figure}
}

{
Migrants are more likely to move to destinations with a significant diaspora, not necessarily places with a large population. However, places with a considerable diaspora also tend to have a large population, so the gravity model works relatively well in those limited cases. Nevertheless, the gravity model fails to capture details at small geographical scales. We compare the results of our model with a gravity model at the neighbourhood level. Vienna is divided into 23 districts (or ``Bezirke''). They are numbered ``outwards'', so the first district is the city centre, and the 20th{--}23rd districts are suburbs. The districts tend to be highly heterogeneous regarding demographic and income compositions. Vienna's 10th district is known for being highly multicultural, attracting nearly 8.3 times more people from Serbia than from Germany. In contrast, the 7th district (known for being the trendy shopping district) attracted 1.4 times more people from Germany than Serbia (Figure \ref{ViennaFig}). There are many reasons why Germans are more likely to move to one neighbourhood in Vienna and Serbians to a different one, but similar reasons have applied to previous migrants. Similar patterns are observed within other countries of origin and city neighbourhoods. 

\begin{figure}[ht] \centering
\begin{center}
\includegraphics[width=0.65\textwidth]{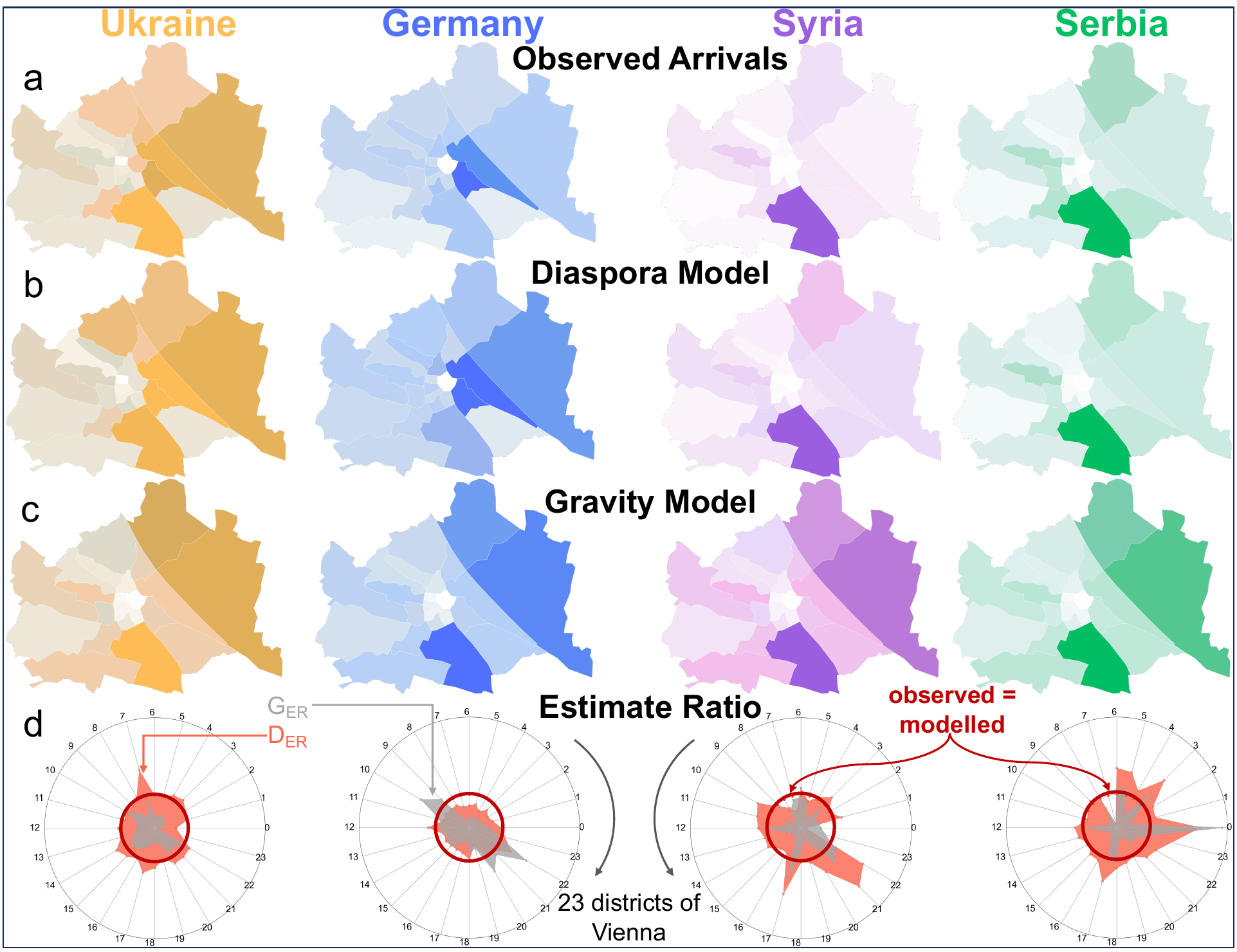}
\caption{\textbf{Vienna model results.}
 a - Heat map of Vienna of the observed arrivals in Austria for the four top diasporas in Austria. b - Heat map of the diaspora model estimates in Vienna. c -  Heat map of the gravity model estimates in Vienna. d - Spider plots of the top four diasporas in Austria, where each section is one of Vienna's 23 districts. The ratio between the modelled and the observed arrivals - estimate ratio - are displayed for each district for both the gravity model in grey ($G_{ER}$) and the diaspora model in pink ($D_{ER}$). The inner circle (red) is when the observed and the modelled arrivals are equal. When the polygons are smaller than the circle, the model underestimates the number of migrants but overestimates that number when it is bigger.}
\label{ViennaFig}
\end{center}
\end{figure}
}

\subsection*{Migration flows in the US}

{
Homophilic migration is also observed in the US. International migrants to the US form a stable process with repeated geographical patterns. Based on the number of arrivals from different parts of the world to a metropolitan area in the US, we estimate the arrivals in the next year (more details in the Methods). The results are also repeated patterns. People from South America, for example, are four times more likely to move to Miami than to Houston. However, the opposite applies to people from Central America, who move more frequently to Houston instead  (Figure \ref{UsMigration}). This phenomenon has persisted for years, although both metropolitan areas have roughly the same population and are at a similar distance to both origins. 

\begin{figure}[ht] \centering
\begin{center}
\includegraphics[width=0.75\textwidth]{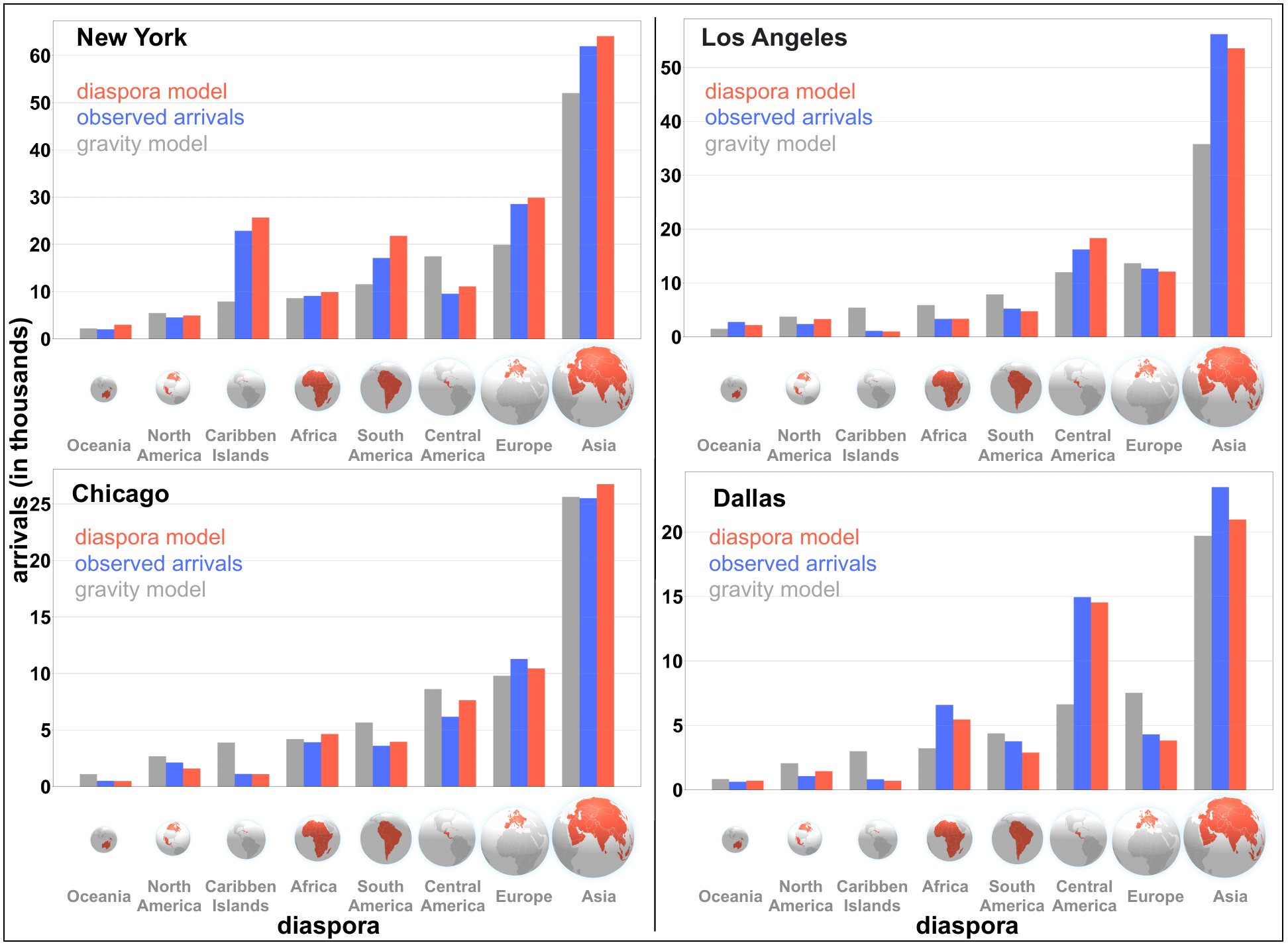}
\caption{\textbf{Top USA metropolitan areas.} Results of the arrival flows of the top four metropolitan areas in the US: New York, Los Angeles, Chicago and Dallas. We plot the diaspora model estimates (red), the observed flows (blue), and the gravity model estimates (grey) for eight estimated diasporas. The diasporas are ranked according to their total arrival flow in the US in 2019. The smallest diaspora is from Oceania, with around 110,000 individuals, while the largest is from Asia, with more than 25 million migrants.}
\label{UsMigration}
\end{center}
\end{figure}
}

{
Migrants do not necessarily choose large or closer cities as their destination. For example, people from Africa are more likely to move to Washington DC than they are to move to New York City, even when New York City has three times more population. On the contrary, people from Europe are two times more likely to move to New York City than to Washington, DC. The diaspora model can distinguish previous migrant populations and their assortativity, which allows us to estimate the inflow of migrants with high precision. 
}

{
We test the diaspora model and compare it with the gravity model. For example, the diaspora model predicts 91,694 migrants to the Miami metropolitan area from all regions considered, but the gravity model only predicts around 39,492 migrants (SM D). However, we observed around 92,500 migrants. This underestimation from the gravity model is persistent in large metropolitan areas such as Washington, New York, Houston and San Francisco, where there is already a significant migrant population acting as a pulling force. 
}

\section{Discussion}

{
Understanding population dynamics and the impact of the arriving migrants into a country is crucial to planning the provision of services and integrating people into the hosting society. Due to conflict, disasters, and a demographic expansion in the countries of origin, combined with the decline in the birth rates in other parts, the share of international migrants will keep increasing in the upcoming decades \cite{UNDesa}. Thus, accurately modelling migration flows will enhance how migration is managed across countries. 
}

{
We distinguish two components of migration: intensity, which captures the number of migrants, and assortativity, which captures their destination once they have decided to move to a new country. For some country of origin, the intensity is modelled as a homogeneous process, meaning that a similar number of arrivals is expected daily. Although there are minor fluctuations (fewer arrivals during the weekend, for example), and there may be some seasonality and other fluctuations, the overall perspective is that the number of arrivals is roughly stable. We show that the size of the diaspora in the country can approximate the intensity rate. Then, the assortativity captures how destinations with big diasporas attract more people from the same country of origin. Our model uses the diaspora's size as the only input and explains migration at small geographical units, such as neighbourhoods. It is impossible to know why many people have moved to a place, but those reasons persist over time and may apply to others. The principle is that most reasons for moving to a new neighbourhood remain and keep attracting similar people. Adding the arrivals at the neighbourhood level gives the estimate at the city level, which also may be combined to obtain an estimate at the state or country level. Thus, diaspora size can be a unique factor in predicting where migrants will move. 
}

{
Our model does not predict migration shocks resulting, for example, from a crisis, as they fall outside the scope of its predictive abilities. However, dividing migration into intensity and assortativity enables modelling migration shocks by altering the intensity of arrivals. In the case of a shock elsewhere, the migration intensity will shift to an unknown value, but the assortativity will still explain where most people will be inclined to move. Nevertheless, when the arrival rate of migrants going through a crisis attains stable features - after a certain period, for example, in Syria and Ukraine- we can alter the intensity of arrivals and predict their expected assortativity.  
}

{
People arriving from different countries go to specific neighbourhoods, so segregation is one of the unintended consequences of this process. Migrants do not necessarily seek to be surrounded by their diaspora, but frequently they are. The diaspora model of migration highlights one of the biggest challenges of migration. Minorities concentrate, fostering fewer interactions with others. This process has important implications, as integrating foreigners into national life is complicated when migrants form segregated communities. Our model helps illuminate the mechanisms when migrants choose their destination and guide policies for designing more inclusive and integrated societies. Governments and international organisations must support local authorities and implement strategies to improve migrant inclusion in urban areas \cite{landau2020local}. 
}

\section{Methods}

\subsection*{Data} 

{
Arrivals and diaspora size for Austria were provided by the Federal Ministry of Interior of Austria, Bundesministerium für Inneres ``BMI''. The data includes all individuals in Austria who register their residence through the mandatory registration form called the ``Meldezettel''. We also have data on asylum seekers of all stages, whether seeking asylum, approved or rejected, and displaced migrants, for example, due to the Russian invasion of Ukraine. Our data does not cover short-term visitors, for example, tourists who are not obligated to register. In addition, we cannot quantify or detect undocumented migrants. Thus, they are not included in our analysis.

Data for the US was obtained from the American Community Survey - Census data \cite{USCensus}. The data contains the resident population and an estimate of the number of arrivals to each metropolitan area (391 areas in total). The data includes the person's residence the year before the survey but not previous years, so repeat migration is not observed. International arrivals are grouped into eight categories: four in America (North America, Central America, South America and the Caribbean) and Africa, Asia, Europe, and Oceania. Data is available in yearly intervals aggregated in five-year periods, from 2009{--}2013 to the 2015{--}2019 surveys.
}

\subsection*{Constructing a diaspora model for the migration}

{
We decompose the migration process with two hierarchical components: the migration intensity (related to the number of migrants from some country) and the assortativity (related to the destination). We use a two-step hierarchical model to consider the two steps separately. This method is frequently used in other domains, where a random variable is modelled by considering ordered steps \cite{kery2010hierarchical}.
}

\subsubsection*{Modelling the migration intensity}

{
We start with the number of migrants and assume it follows a Poisson distribution such that:
\begin{equation} \label{ModelArrivals}
    M_i(t) \,\,\, \sim \,\,\, \text{Pois}( \lambda_i t),
\end{equation}
where $M_i(t)$ is the flow of migrants for a period of $t$ days, from the country of origin $i$, where $i = 1, 2, \dots, \mu$ and with a daily rate $\lambda_i$. The Poisson distribution is frequently used to model a variable that results from a counting process, such as migrants \cite{PrietoColombiaMobility}. The distribution depends on a single rate, and it is used for ignoring short-term fluctuations and looking only at the more general pattern. The expected number of migrants until day $t$ is $\lambda_i t$, an expression for the cumulative number of arrivals from country $i$. 
}

{
First, we test if a uniform rate works during $n$ days for estimating the arrival rate from different countries. The error term for day $i$ gives $e_i(t) = M_i(t) - \lambda_i t$. The sum of square errors over the observed days gives $f(\lambda_i) = \sum_{t = 0}^n e_i(t)^2$. By setting $f'(\lambda_i) = 0$ we obtain that 
\begin{equation} \label{LambdaStar}
\lambda_i^\star = \frac{6 \sum_{t = 0}^n t M_i(t)}{n(n+1)(2n+1)}.
\end{equation}
Since $f(\lambda)$ is a continuous function and $f''(\lambda_i) = n(n+1)(2n+1)/3 > 0$, then the value $\lambda_i^\star$ minimises the error. For a sufficiently large number of days and arrival rate, a Normal approximation to the Poisson distribution may be used to obtain the corresponding confidence interval, given by $[\lambda_i t - \theta \sqrt{\lambda_i t}, \lambda_i t + \theta \sqrt{\lambda_i t}]$. For a lower rate (or for fewer days), a Monte Carlo method may also be used to obtain plausible departures. For the top migrant countries to Austria, the estimated arrivals are approximated by the constant rate (Figure \ref{IntensityConeTest}). 

\begin{figure}[ht] \centering
\includegraphics[width=0.85\textwidth]{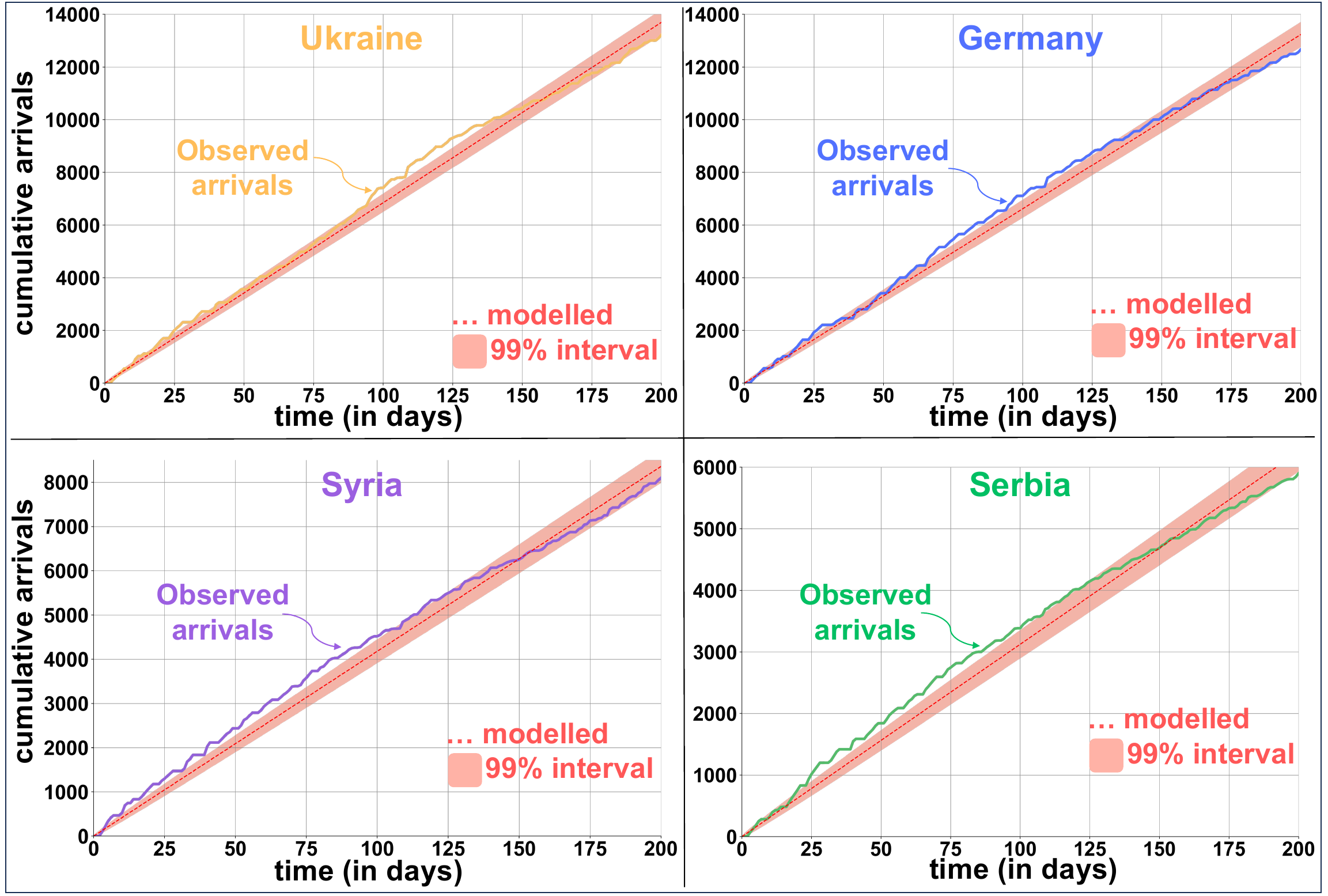}
\caption{\textbf{Estimating the pull rate ($\lambda_i$).} The intensity of migration for country $i$ gives the daily arrival rate for that country $\lambda_i$. For $t$ days, the expected number of arrivals is $\lambda_i t$, plotted as a dashed line for each country. For a sufficiently large rate and number of days, a Normal approximation gives a 99\% confidence interval, plotted for each country as the shaded triangle (with $\theta = 4$). The observed number of arrivals falls within the shaded triangle, so we do not reject a constant arrival rate for those countries of origin. }
\label{IntensityConeTest}
\end{figure}
}

{
There are some fluctuations in the daily number of arrivals. For example, very few people arrive during the weekends. However, a fixed arrival rate works well for modelling the daily arrival of migrants and enables us to ignore minor fluctuations. Equation \ref{LambdaStar} may be used to estimate the daily arrival rate for migrants from different countries. Although Equation \ref{LambdaStar} estimates the arrival rate, we then aim to approximate the rate based on the size of the country's diaspora. The motivation is to model whether a bigger diaspora results in more arrivals (as observed in Figure \ref{Flows}). We take the data for all countries, considering that
\begin{equation} \label{RegressionDiaspora}
M_i(t) = \rho R_i t,
\end{equation}
where $R_i$ is the size of the diaspora from country $i$, and $\rho > 0 $ is a fixed pull rate for all countries of origin and depends only on the arrivals and existing diasporas at the destination. Thus, we assume that for a given country, the flow depends on the size of the diaspora $R_i$ and some fixed pull rate $\rho$. Following the same logic, we obtain the value of $\rho$ by minimising the errors such that the error $e_i = M_i - \rho  R_i t$. The sum of the squared errors over the observed days gives $g(\rho) = \sum_i e_i^2$ By setting $g' =0$, we obtain that 
\begin{equation}\label{eq:Pull_Destination}
    \rho^\star = \frac{\sum_{i=1}^{\mu} M_i R_i}{t \sum_{i=1}^{\mu} R_i^2},
\end{equation}
}
an estimate for the pull rate that depends on the arrivals and diaspora of all countries. The second derivative of $g$ is $2 \sum_{i = 1}^\mu R_i^2 t^2 >0$, so the value of $\rho^\star$ minimises the sum of the squared differences. Equation \ref{eq:Pull_Destination} depends on the size of the diaspora and the arrivals over $\mu$ countries, unlike Equation \ref{LambdaStar}, which depends on the daily arrivals for a single country.

{
In the case of Austria, we obtain that $\rho_{Aus} = 3.29 \times 10^{-4}$. Then, we can use the diaspora size and the estimated pull rate to express the arrival rate for country $i$ as 
\begin{equation} \label{DiasporaPull}
    \lambda_i = \rho R_i.
\end{equation}

Equation \ref{DiasporaPull} gives the arrival rate from county $i$ considering the diaspora size of that country and a pull rate $\rho$ that applies to all countries equally. This method overestimates arrivals from some countries (for example, Serbia and Turkey) and underestimates others (for example, Romania and Germany) and, in general, results in higher error than Equation \ref{LambdaStar} (SM C). However, it gives an alternative expression to estimate rates that do not depend on the data for the daily arrivals to that country. The obtained value of $\rho_{Aus} = 3.29 \times 10^{-4}$ reflects that one person is expected to arrive daily for every 3,031 individuals from any diaspora. For example, in Austria, there are around 42,580 people from Poland, so 14 people are expected each day, or 2,800 people during the 200 days of observation. The observed number of arrivals from Poland during that period was 2,787 migrants (SM A). 
}

\subsubsection*{Modelling the migration assortativity}

{
Once the number of arrivals is known, we model their conditional destination in that country \cite{ZIPMigration, cohen2008international, lee2018hierarchical}. We assume that once $M_i(t) = m$ persons arrive, they decide to reside in a particular location $j$ - for example, a city in the destination country - depending on the size of the diaspora in the location $j$. Thus, we assume that the probability of a person from country $i$ moving to location $j$ follows:
\begin{equation*}
    \pi_{ij} = R_{ij} / R_i,
\end{equation*}
where $R_{ij}$ is the diaspora from country $i$ in destination $j$. The diaspora is such that $R_i = \sum_j R_{ij}$ is the overall size of the diaspora from country
$i$. For example, if location $j$ has 10\% of the diaspora from $i$, we assume that the probability that a migrant moves to $j$ among the $\nu$ destinations is also 10\%. Destinations with bigger diaspora attract more migrants. The process is modelled as a Multinomial distribution:
\begin{equation*}
    M_{ij}(t) | M_i(t) = m \,\,\, \sim \,\,\, \text{Mult}(m, \bar{\pi_i}),
\end{equation*}
where $M_{ij}(t)$ are the arrivals from $i$ in location $j$ and $\bar{\pi_i} = (\pi_{i1}, \pi_{i2}, \dots, \pi_{i\nu})$ is the vector with entries $\pi_{ij}$ corresponding to the probability of choosing $j$ as their destination, where $\sum_k \pi_{ik} = 1$. A Multinomial distribution, conditional on a Poisson distribution, also follows a Poisson distribution with combined rates of arrivals and success \cite{PrietoColombiaMobility}. Therefore, arrivals from country $i$ to location $j$ follow $M_{ij}(t) \,\,\, \sim \,\,\, \text{Pois}( \pi_{ij} \lambda_i t)$. 
}

{
If the daily rate of arrivals is known, then Equation \ref{LambdaStar} gives an estimate of the arrivals at a granular level. If the diaspora size from other countries is known, then Equation \ref{DiasporaPull} gives an estimate that depends only on the diaspora size from country $i$ in destination $j$. Combining both, arrivals from country $i$ to location $j$ follow 
\begin{equation}
    M_{ij}(t) \,\,\, \sim \,\,\, \text{Pois}( R_{ij} \lambda t).
\end{equation}\label{eq.main}
}

{
Other migration models focus only on the assortativity of individuals, meaning that they only differentiate the likelihood of moving between destinations but ignore the number of people moving. However, migration intensity is more relevant than assortativity. Modelling how many people will arrive in a country should be the first explicit component of migration models. Arrivals can be modelled with a constant rate so that they can be predicted within a reasonable period. Further, the arrival rate is strongly linked to their diaspora size. We estimate how many people will move to a country if we observe how many people already live there. The second element, related to assortativity, has been captured by size, job offerings and others. Yet, diaspora size explains assortativity more accurately and with smaller geographical units than metropolitan areas. Because large cities tend to have a high diaspora share, the gravity model works relatively well to capture a general trend. However, the gravity model fails at smaller geographical units. According to the gravity model, two neighbourhoods of similar size are equally attractive to migrants, but this is never the case. Ethnic neighbourhoods, such as ``Chinatowns'', are part of the cultural landscape in most cities. 
}

\subsection*{Inferring diaspora size and subsequent migrations} 

{
In the case of international migration to metropolitan areas in the US, census data estimates the yearly arrivals between 2009 and 2019 \cite{USCensus}. It is disaggregated for eight regions of origin (Africa, Asia, the Caribbean, Central America, Europe, Northern America, Oceania, and South America). It gives the number of arrivals to 391 metropolitan areas and the countryside. We use the number of arrivals in one year to estimate arrivals in the subsequent one. Let $A_{ij}(t)$ be the number of arrivals from region $i$ to metropolitan area $j$. We assume that the arrivals in that year result from some fixed pull rate $\lambda$ and an unknown diaspora size $D_{ij}$. Then, the expected number of arrivals for the next year follows a Poisson distribution with rate $\lambda D_{ij} = A_{ij}$ for some value of $\lambda > 0$. Thus, the expected number of arrivals in one year is the observed number during the previous year. 
}

{
We compare the estimated number of arrivals $E_{ij}$ to the observed number $A_{ij}$ in 2019. At the regional level, for example, the model gives that 131,478 arrivals are expected from Africa, and between 130,767 and 132,188 arrivals are expected. The observed number of migrants from Africa in 2019 was 131,943. For other regions, the estimated inflow is within a 3.5\% difference from the observed number of arrivals, except for the case of South America. The yearly number of migrants from South America to the US has nearly doubled in seven years. This increasing intensity in the annual number of migrants would be better captured considering the observed trend (more details in SM D).
}

\section{Supplementary Materials}

\subsection*{A - Data description and observation}

{
The data corresponding to arrivals to Austria is provided in the \textbf{Complex Effects of Migration Patterns on Supply Capacities} project. Access to the data is restricted for security and privacy reasons, and only authorised researchers can view the data. For each person, the data includes their country of origin and the neighbourhood in which they have a registered residence. The data gives the location of the diaspora for each country at the postal code level. Data before November 26, 2022 (referred to as ``December'' in the manuscript) does not identify each arrival date. It gives the age, gender, residential status, and country of origin of $R = 1,466,113$ migrants in Austria. For 200 days, the majority of arrivals to the country were captured at the moment when a person registered their residence through a registration form known as the ``Meldezettel'' with the Bundesministerium für Inneres (Federal Ministry of the Interior) when they apply for some form of residence permit in the country. Visits planned for shorter periods (tourism) do not require registering and are not counted.  
Legally, migrants in Austria are classified according to their residence status. For example, migrants who plan to stay in the country for less than six months are classified as foreigners, but if they stay longer, they are classified as settled migrants with a residence permit. On the other hand, refugees fall under a different classification depending on their stage, for example, seeking asylum, approved or rejected. As of 14 June 2023, refugees cover only 13.72\% in our data (Table \ref{Migrant Classfications}). We apply the same analysis to all classifications. 

\begin{table}[ht!]
\centering
\begin{tabular}{l r r} 
 \hline
 \textbf{Migration Status}  & \textbf{Percentage} \\
 \hline
Settled migrant with a residence permit   & 54.30 \% \\
Foreigner      &  31.76\% \\
Entitled to asylum     & 10.78\% \\
Eligible for subsidiary protection      & 1.70\% \\
Approved asylum seeker   &  0.63\% \\
Displaced person      & 0.53\% \\
Humanitarian residence permit     &  0.08\% \\
Asylum seeker &  0.01\% \\
\hline
Other classifications & 0.21\% \\
 \hline
\end{tabular}
\caption{Migrants residential status as of 14 June 2023}
\label{Migrant Classfications}
\end{table}
}

{
The data contains information describing the nationality and residence of 1.46 million people from 192 countries. As of November 26, 2022, Austria has 1,542,349 registered migrants from 192 countries. Around 95\% disclosed the main addresses and are considered here. We analyse the arrivals for 263 days divided into two parts: 200 days to train and 63 days to test. Within the period of analysis that considers 200 days, there were $A = 111,244$ arrivals to the country, mainly from Ukraine, Romania, Germany and Syria. As of 14 June 2023 (after 200 days), around 75\% of arrivals are from 15 countries (Table \ref{ArrivalsTableSupp}). 

\begin{table}[ht!]
\centering
\begin{tabular}{l r r r} 
 \hline
 \textbf{Country of origin} & \textbf{Diaspora} & \textbf{Arrivals} & \textbf{Arrivals \%}\\
 \hline
Ukraine &   76,577 &  12,054 & 10.84\% \\
Romania &  102,314 &  11,065 &  9.95\% \\
Germany  &  115,913 &  10,064 &  9.05\% \\
Syria &   82,746 &   8,030 &  7.22\% \\
Hungary &   65,269 &   7,794 &  7.01\% \\
Croatia &   89,296 &   5,177 &  4.65\%\\
Turkey &  135,205 &   4,926 &  4.43\% \\
Serbia &  145,348 &   4,536 &  4.08\% \\
Bosnia and Herzegovina &  109,478 &   3,462 &  3.11\% \\
Afghanistan &   52,821 &   3,086 &  2.77\% \\
Bulgaria &   27,474 &   2,934 &  2.64\% \\
Poland &   42,580 &   2,787 &  2.51\% \\
Slovakia &   32,363 &   2,650 &  2.38\% \\
Russia &   41,013 &   2,508 &  2.25\% \\
Italy &   21,581 &   2,264 &  2.04\% \\
\hline
Other countries & 326,135& 27,907& 25.09\%\\
 \hline
 \textbf{Total}  &   1,466,113& 111,244& \textunderscore\\
 \hline
\end{tabular}
\caption{The top countries of origin in Austria, in descending order of arrivals within the observation period of 200 days. Only migrants with registered main addresses. We list countries with arrivals percentages above 2\%. The diaspora is the country's pre-existing population size - before November 26, 2022. }
\label{ArrivalsTableSupp}
\end{table}

}

{
We test whether a uniform daily arrival explains the observed number of migrants from the top countries of origin. A uniform daily arrival is not rejected for the top 12 countries. The observed arrivals fall within the modelled intervals (Figure \ref{IASupp}). We also test whether postal codes with a larger diaspora attracted more migrants. 
\begin{figure}[ht!] 
\includegraphics[width=0.85\textwidth]{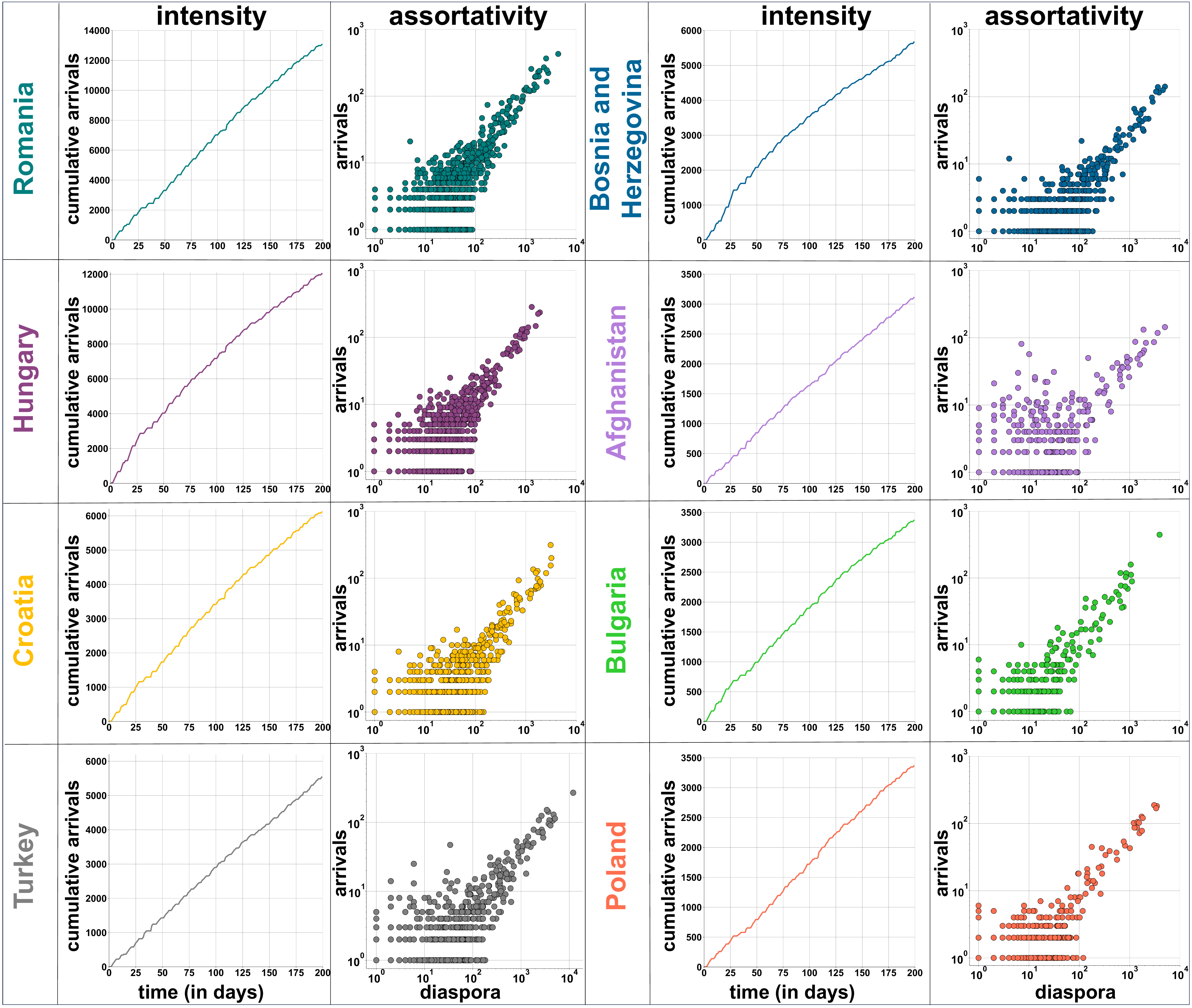}
\caption{\textbf{Intensity and assortativity of the top Diasporas.} The intensity of Arrivals in the 200 days of observation of the top eight diasporas in Austria (left). The assortativity of the arrivals of the top eight diasporas (right).}
\label{IASupp}
\end{figure}
}

\subsection*{B - Modelling intensity and assortativity}

{
Let $M_i(t)$ be the number of arrivals from country $i$ since time $t = 0$. We assume that $M_i(t) \,\,\, \sim \,\,\, \text{Pois}( \lambda_i t)$, so the expected number of arrivals during $t$ days is $\lambda_i t$. The Poisson distribution is frequently used to model discrete events (such as the number of arrivals) since it allows overlooking small perturbations or fluctuations and focuses on the more general picture, the daily arrivals. It depends on a single parameter, $\lambda_i$, known as the (daily) rate, which is the expected number of arrivals per day. 

Once a person decides to move to some country, they decide on a specific location, which can be as general as states or provinces or as particular as neighbourhoods. The person chooses location $j$ with probability $\pi_{ij}$. The destination, conditional on observing $m$ arrivals, can be considered a Multinomial distribution with $\nu$ options. The vector $\bar{\pi_i} = (\pi_{i1}, \pi_{i2}, \dots, \pi_{in})$ captures the destination preferences for people from origin $i$. In particular, the decision of moving to destination $k$, with $k \in 1, 2, \dots, n$, is a Binomial distribution (with a probability of success $\pi_{ij}$ and with a probability of failure $1-\pi_{ij}$). Thus, the number of arrivals to destination $k$, conditional on observing $m$ arrivals, 
\begin{equation}
M_{ik}(t) | M_i(t) = m  \,\,\, \sim \,\,\,\text{Bin}(m, \pi_{ik}).
\end{equation}
It is possible to show that a Binomial distribution, conditional on a Poisson distribution, is also a Poisson distribution \cite{PrietoColombiaMobility}. It has a rate $\lambda_i \pi_{ik}$, which is the same rate but discounted by the probability $\pi_{ik}$. Thus, arrivals to destination $k$ are $M_{ik}(t) \,\,\, \sim \,\,\, \text{Pois}( \lambda_i \pi_{ik} t)$. Modelling intensity and assortativity separately enables the disentangling of the process with a minimal set of parameters.

{
The assortativity of our model of migration estimates that people from $i$ move to location $j$ with probability $\pi_{ij}  = R_{ij} / R_i$, where $R_{ij}$ is the diaspora size of country $i$ in location $j$, and $R_i = \sum_j R_{ij}$ is the total diaspora. After some period, $t$, the new diaspora will have size $R_{ij}'(1 + b - d + i - o + \lambda)$, where $b$ is the birth rate, $d$ is the death rate, $i$ is the inflow due to internal movements, $o$ is the outflow due to internal movements, and $\lambda$ corresponds to the new arrivals. Assuming that the impact of internal migration of the diaspora is negligible (meaning that $i \approx o$), we get that $R_{ij}'(1 + b - d + \lambda)$. Further, assuming that the birth and death rates are the same for all the diasporas, we get that $R_i' = \sum_j R_{ij}'(1 + \lambda) = R_{i}'(1 + \lambda)$, so the total diaspora also changes size due to the arrival of people. Then, the assortativity impact, after some period is $\pi_{ij}' = R_{ij}' / R_{i}' = \pi_{ij}$, so it remains unchanged. Thus, the model conserves the distribution of the diaspora across destinations after the arrival of people is considered.
}
}

\subsection*{C - Model comparison}

In this section, we compare our model with the gravity model. The gravity model is one of the most prominent ways in which social mobility is analysed. The gravity model captures the impact of size at the origin and destination countries and their distance \cite{Gravity, GravityModel, MigrationLaws1885, stouffer1940intervening, jung2008gravity}. Gravity has been used, for example, to model trade between countries and cultural distances or frictions between distinct locations \cite{Gravity, GravityModel, barbosa2018human}. The gravity model, however, does not quantify the intensity of migration but gives only a description of the assortativity. One of the most significant drawbacks of the gravity model is that it does not consider any temporal dimension, so it only ranks destinations depending on their size. Unfortunately, the gravity model does not provide the expected arrivals of migrants or an analogy to our diaspora pull rate; thus, we do not include it in the intensity error calculations.

\subsubsection*{Intensity}

{
To assess the error in the expected arrivals of migrants, we use the 200 days of observations to construct a daily pull rate for every country $\lambda_i$ (Equation \ref{LambdaStar}) and predict the arrivals in the next nine weeks (63 days since 14 June 2023), we choose to have a time window in weeks instead of months because migration patterns and data registration goes through a weekly cycle. We compare our estimate with the actual observed arrivals and find that our model can predict the observed arrivals with a margin of $\pm 0.17$ arrival per country per day (Table \ref{Tab:Intensity_Err}). Additionally, we compute the daily pull rate of Austria $\rho_{Aus}$ (Equation \ref{eq:Pull_Destination}) and estimate the arrivals for all countries and find that using this method, our model can predict the observed arrivals with a margin of $\pm 0.32$ arrival per country per day. Thus, using $\rho_{Aus}$, we get almost twice the error. However, we can rely on fewer data points.
\begin{table}[ht!]
\centering
\begin{tabular}{lrrrrrr}
\hline
\textbf{Country} &  \textbf{$\lambda_i$} &       \textbf{Arrivals}& \textbf{Arr($\lambda_i$)} & \textbf{Arr($\rho_{Aus}$)} &\textbf{Err($\lambda_i$)} &\textbf{Err($\rho_{Aus}$)}\\
\hline
Ukraine &  68.48 &   3,106 & 4,314 &  1,842 & 1,459,264 & 1,597,696 \\
     Hungary &  65.46 &   3,264 & 4,124 &  1,519 &  739,600 & 3,045,025 \\
     Syria &  41.79 &   3,309 & 2,633 &  1,887 &  456,976 & 2,022,084 \\
     Germany &  66.18 &   3,579 & 4,170 &  2,618 &  349,281 &  923,521 \\
     Romania &  67.44 &   3,713 & 4,249 &  2,356 &  287,296 & 1,841,449 \\
     Slovakia &  26.59 &   1,180 & 1,675 &   728 &  245,025 &  204,304 \\
     Bosnia and Herzegovina &  31.05 &   1,462 & 1,956 &  2,347 &  24,4036 &  783,225 \\
     Seriba &  31.25 &   1,561 & 1,969 &  3,115 &  166,464 & 2,414,916 \\
     Bulgaria &  17.83 &    796 & 1,123 &   632 &  106,929 &   26,896 \\
     Italy &  17.30 &    833 & 1,090 &   496 &   66,049&  113,569 \\
\hline
All Countries & \textunderscore & 39,360 & 44,559 & 32,777 & 4,384,765& 15,312,703\\
\hline 
$\sqrt{\sum (E_r)^2}$ &  \textunderscore & \textunderscore &  \textunderscore &\textunderscore  & $\mp 0.17$ & $\mp 0.32$ \\
\hline
\end{tabular}
\caption{Top 10 errors comparisons of the observed and modelled arrivals where Arrivals are the observed arrivals, $\lambda_i$ is the daily pull rate of every country, $\rho_{Aus}$ is the daily pull rate of Austria, Arr($\lambda_i$) are the modelled arrivals using $\lambda_i$, Arr($\rho_{Aus}$) are the modelled arrivals using $\rho_{Aus}$, Err($\lambda_i$) and Err($\rho_{Aus}$) are the squared error of Arr($\lambda_i$) and Arr($\rho_{Aus}$) respectively. The errors are ranked according to the squared error of $\lambda_i$ (Err($\lambda_i$))in descending order for nine weeks (63 days). $\sqrt{\sum(E_r)^2}$ is the square root of the sum of the squared error averaged over 63 days of observation and 192 countries.}
\label{Tab:Intensity_Err}
\end{table}
}

\subsubsection*{Assortativity}

{
For a fixed period, a country of origin $i$ and destination $j$, we have modelled the flow $D_{ij}$ and compared it to the observed flow $M_{ij}$. We compute the mean square error as:
\begin{equation}\label{eq:TotalErr}
    E_r = \sum_{i, j} \frac{(D_{ij} - M_{ij})^2}{\mu \nu},
\end{equation} 
where $\mu$ and $\nu$ are all the possible origins and destinations. The mean squared error can be used to compare distinct models, where a smaller error means better performance. 
}

{
The gravity model assumes that destination $j$ with population $P_j$ attracts population depending on its size, so we consider its assortativity as $\pi^{g}_{ij} = f(P_j, D_{ij})$ for some function $f$ that takes the size of the destination and the distance between origin and destination. We construct a gravity model $\mathcal{G}$ such that once a person has decided to move to a country, they choose their destination depending on its size. Thus, we also consider that the destination is picked as a Multinomial distribution depending on its size. Formally, we assume that once $m$ people have moved from $i$, they will move to $j$ depending on its size, so $\pi^{\mathcal{G}}_{ij} = P_{j}^\alpha / \sum_j P_{j}^\alpha$, for some parameter $\alpha \geq 0$. We compare the diaspora and gravity models by comparing the mean square error (Figure \ref{ModelInfog}). 

\begin{figure}[ht!] \centering
\includegraphics[width=0.8\textwidth]{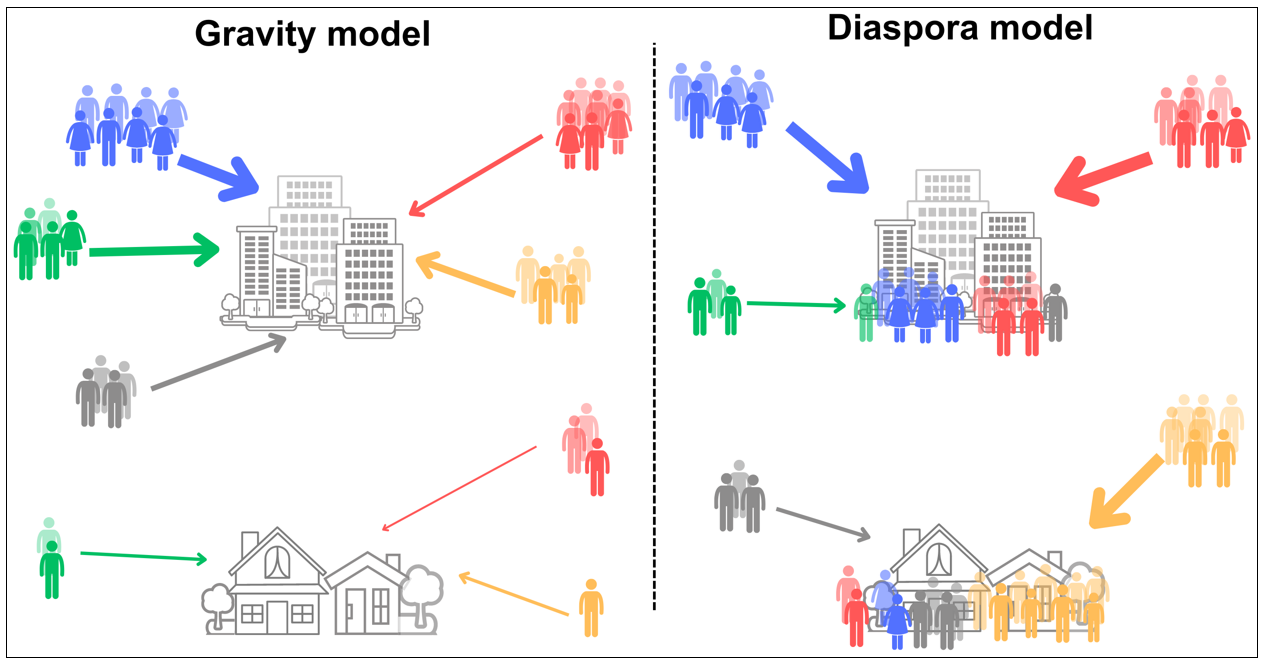}
\caption{ \textbf{Gravity and Diaspora model descriptions.}. We divide migration into two components: intensity (related to the arrival of individuals) and assortativity (related to where migrants decide to go). The diaspora model of migration uses the size of the pre-existing population of a certain diaspora. The gravity model uses the total pre-existing population without accounting for diasporas and individual differences.}
\label{ModelInfog}
\end{figure}
}

{
We compare the diaspora and the gravity model by analysing only their assortativity. That is, we assume $m$ arrivals to some destination and distribute them depending on the assortativity of the model considered. We consider the sum of the squared error terms of each model (where a smaller error means that the model describes more accurately the assortativity of the arrivals). The difference between the error terms of the gravity and the diaspora models is enormous, particularly for the countries with the highest number of arrivals. For example, for arrivals from Germany, the gravity model has a squared error of 280.97, but the diaspora model has a squared error of 55.60. 
}

{
We use the error equation outlined in Equation \ref{eq:TotalErr} to calculate the error of each origin country for all destinations. We average the error over 100 simulations for both models and get the average per country (Table \ref{tab:Err_detailed_Countries}). In total, there are 192 countries and 2,221 possible destinations. 

\begin{table}[h!]
   \centering
  \begin{tabular}{lrr} 
  \hline
\textbf{Country} &      \textbf{Diaspora Model} &       \textbf{Gravity Model}\\
{} & \textbf{square error} & \textbf{square error}\\
\hline
Syria &   67.39 &  389.42 \\
Germany &   55.60 &  280.97 \\
Ukraine &   56.78 &  224.34 \\
Morocco &  185.32 &  184.17 \\
Romania &   39.36 &  150.83 \\
Seriba &   19.97 &  123.61 \\
Bulgaria &    8.59 &   91.35 \\
Croatia &   21.17 &   75.07 \\
Hungary &   20.59 &   44.21 \\
Poland &    8.19 &   44.14 \\
\hline
All Countries & 658.45 & 1,875.97\\
\hline 
$E_r$ &  3.42 & 9.77\\
\hline
\end{tabular}
\caption{Error Comparison between the diaspora and the gravity model. We list the top 10 countries of origin for all postal codes, sorted in descending order according to the gravity model such that the Syria diaspora has the biggest gravity model error and the Poland diaspora has the lowest. $E_r$  is the mean squared error.}
\label{tab:Err_detailed_Countries}
\end{table}
}

{
A crucial aspect of migration models is considering different geographic levels. For example, detecting the number of arrivals at the province level is critical since some provisions are frequently managed at that level (such as health or education). However, in smaller units such as cities and neighbourhoods, forecasting the number of migrants plays a critical role. One of the most significant weaknesses of the gravity model is that it cannot predict migration at the neighbourhood level. The gravity model has a squared error of 4,925.15 when we look at the arrivals to the 10th district of Vienna (Favoriten), but the diaspora model has a squared error of 500.76. Results show that the mean square error is 3.42 for the diaspora model but 9.77 for the gravity model. Thus, the average error of each destination for all countries is nearly three times bigger for the gravity model compared to the diaspora model (Table \ref{tab:Err_detailed postal}).

\begin{table}[ht!]
\centering
\begin{tabular}{lrr}
\hline
\textbf{Postal Code} &           \textbf{Diaspora Model} &       \textbf{Gravity Model}\\
{} & \textbf{squared error} & \textbf{squared error}\\
\hline
1100 &  500.76 & 4,925.15 \\
6020 &  393.12 & 2,288.01 \\
8020 &  211.23 & 1,104.15 \\
8055 &  912.89 &  950.74 \\
1030 &  138.67 &  859.40 \\
1020 &  110.07 &  806.55 \\
4880 &  764.52 &  788.96 \\
1160 &   97.56 &  754.73 \\
1120 &  161.68 &  618.12 \\
1200 &   90.19 &  488.60 \\
\hline
All Postal Codes & 7,616.84 & 21,700.769\\
\hline 
$E_r$ &  3.42 & 9.77\\
\hline
\end{tabular}
\caption{Top 10 error comparison between the diaspora and the gravity model per postal code for all countries of origin, sorted in descending order according to the gravity model. The postal code 1100 (Favoriten, 9th district in Vienna) has the biggest gravity model error, and the postal code 1200 (Brigittenau, 20th district in Vienna) has the lowest. $E_r$ is the mean squared error.}
\label{tab:Err_detailed postal}
\end{table}
}

\subsection*{D - International migration to the USA} 

{
We conduct our analysis of 387 USA Metropolitan areas - Mets. We exclude movements to the countryside and five areas Metropolitan areas added in the 2019 census. The census data also limits us to only eight diasporas where the migrants' countries of origin are classified: Asia, Europe, Central America, South America, Africa, the Caribbean Islands, North America and Oceania. We use the census data from 2013 to 2018 to estimate the arrivals of our selected Mets in 2019. 
}

{
The gravity model proves insufficient to predict the migration flows with underestimation in big metropolitan areas and overestimation in small metropolitan areas (Figure \ref{Us_AllMets}).

\begin{figure}[ht!] \centering
\begin{center}
\includegraphics[width=0.85\textwidth]{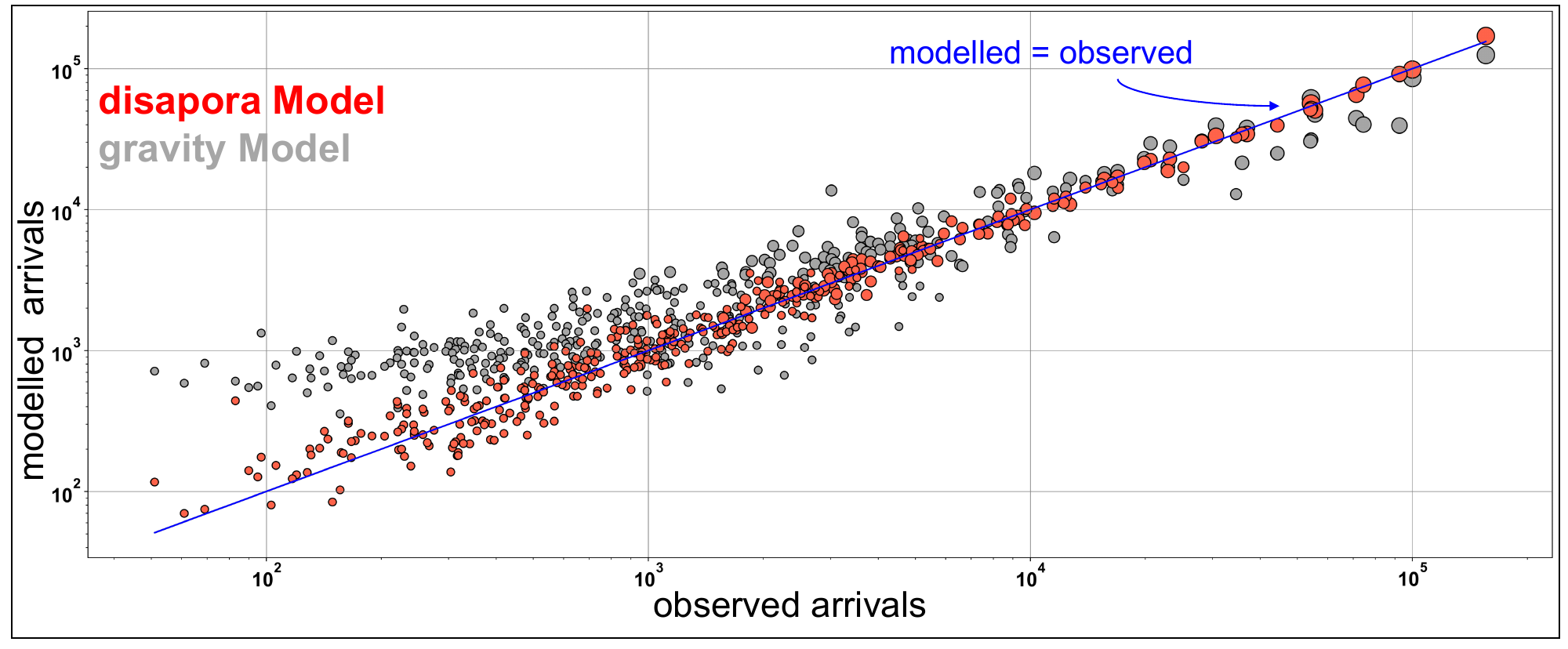}
\caption{\textbf{USA Metropolis.} Results of the arrival flows of all the metropolitan areas 387 in the US. We plot the diaspora model estimates (pink), the gravity model estimates (grey) and the observed flows (blue). The sizes of the observations vary depending on the size of Met.}
\label{Us_AllMets}
\end{center}
\end{figure}
}

{
We use the observed total arrivals in both the diaspora and the gravity model, and we model the assortativity according to Met size in the case of the gravity model and according to average diaspora assortativity in the case of the diaspora model (Table \ref{tab:US_Arrival}).  

\begin{table}[ht!]
\centering
\begin{tabular}{lrrr}
\hline
\textbf{Met Name}  &  \textbf{Arrivals} &      \textbf{Gravity Model} &      \textbf{ Diaspora Model}  \\
\hline
  New York-Newark-Jersey City &    155,722 & 125,025.22 & 170,453.85  \\
     Los Angeles-Long Beach-Anaheim  &     99,989 &  85,882.89 &  98,584.74\\
     Chicago-Naperville-Elgin  &     54,275 &  61,604.83 &  56,617.11 \\
        Dallas-Fort Worth-Arlington &     55,590 &  47,356.62 &  50,547.42 \\
   Houston-The Woodlands-Sugar Land  &     71,303 &  44,491.35 &  65,255.46 \\
Washington-Arlington-Alexandria  &     74,449 &  40,127.31 &  76,847.79 \\
Miami-Fort Lauderdale-Pompano Beach  &     92,500 &  39,492.34 &  91,694.38 \\
Philadelphia-Camden-Wilmington  &     30,614 &  39,414.61 &  33,362.91 \\
   Atlanta-Sandy Springs-Alpharetta  &     36,869 &  37,973.96 &  34,408.14 \\
         Boston-Cambridge-Newton  &     54,319 &  31,350.37 &  52,575.47 \\
\hline
All Mets  &1,823,840 & 1,823,840 & 1,823,840
\end{tabular}
\caption{The observed arrivals of the biggest ten Metropolitan areas in the US in 2019 and their gravity and diaspora model estimates.}
\label{tab:US_Arrival}
\end{table}
}

{
We compare the diaspora model results with the observed arrivals and gravity model results, and we show that the mean squared error of the gravity model is 19.3 times bigger than the diaspora model (Table \ref{tab:USErr}).

\begin{table}[ht!]
\centering
\begin{tabular}{lrr}
\hline
\textbf{Met Name} &         \textbf{Gravity Model} &       \textbf{Diaspora Model}\\
{} & \textbf{squared error} & \textbf{squared error}\\
\hline
  New York-Newark-Jersey City & 942,292,280.17 & 217,027,451.26 \\
     Los Angeles-Long Beach-Anaheim & 198,982,433.19 &   1,971,940.83 \\
     Chicago-Naperville-Elgin & 53,726,416.07 &   5,485,484.20 \\
        Dallas-Fort Worth-Arlington& 67,788,545.62 &  25,427,649.57 \\
   Houston-The Woodlands-Sugar Land & 718,864,340.74 &  36,572,692.26 \\
Washington-Arlington-Alexandria & 1,177,978,170.36 &   5,754,213.28 \\
Miami-Fort Lauderdale-Pompano Beach & 2,809,811,712.71 &    649,017.76 \\
Philadelphia-Camden-Wilmington & 77,450,653.74 &   7,556,509.42 \\
   Atlanta-Sandy Springs-Alpharetta & 1,220,942.69 &   6,055,836.33 \\
         Boston-Cambridge-Newton & 527,558,106.12 &   3,039,895.61 \\
\hline
All Mets   &9,314,104,678.92& 482,227,713.95 \\
\hline 
$E_r$ & 3,008,431.74 & 155,758.30
\end{tabular}
\caption{The squared error of the gravity and the diaspora model in the ten biggest Metropolitan areas, $E_r$ is the mean squared error calculated over all diasporas and Mets.} 
\label{tab:USErr}
\end{table}
}

\section*{Acknowledgements}

We thank Stefan Thurner, Ljubica Nedelkoska and Isabela Rosario Villamil for their comments and insights.

\section*{Competing interests}

The authors declare that they have no competing interests.

\section*{Author's contributions}

RPC designed the study. OA and ED worked with data analysis and visualizations. RS and GH helped with data curation and management. FK, EO and YH analysed the results. All authors wrote the manuscript.

\section*{Funding}

This research is funded by the Federal Ministry of the Interior of Austria (2022-0.392.231)

\section*{Data availability}

The raw and processed data are not available due to privacy laws. The Federal Ministry of the Interior of Austria safeguarded the dataset and made it accessible to our research institution under strict data protection regulations. Researchers must find individual agreements with the Federal Ministry of the Interior of Austria to access this data.

\bibliographystyle{ieeetr}

\end{document}